\documentclass[a4paper,11pt]{article}
\usepackage[latin1]{inputenc}
\usepackage{epsfig}
\usepackage{amssymb}
\usepackage{amsmath}

\author{J. P. Juchem Neto\footnote{Graduate Student, Department of Mathematical Sciences, University of Delaware. E-mail: jp.juchem@yahoo.com.br.}}
\title{Nonlinear Pendulum: A Simple Generalization}
\begin{document}
\maketitle

\begin{abstract}
In this work we solve the nonlinear second order differential equation of the simple pendulum with a general initial angular displacement ($\theta(0)=\theta_0$) and velocity ($\dot{\theta}(0)=\phi_0$), obtaining a closed-form solution in terms of the Jacobi elliptic function $\text{sn}(u,k)$, and of the the incomplete elliptical integral of the first kind $F(\varphi,k)$. Such a problem can be used to introduce concepts like elliptical integrals and functions to advanced undergraduate students, to motivate the use of Computer Algebra Systems to analyze the solutions obtained, and may serve as an exercise to show how to carry out a simple generalization, taking as a starting point the paper of Bel\'endez \emph{et al} \cite{belendez}, where they have considered the standard case $\dot{\theta}(0)=0$.
\end{abstract}

\section{Introduction}
An usual way to present the simple pendulum model in a first course of general physics is to consider a small angular displacement, and in this way to linearize the corresponding second order differential equation \cite{halliday}. This first approximation lead us to conclude that, for small amplitudes, the period of oscillation is independent from the amplitude:
\begin{equation}
\label{periodsmall}
T\approx 2\pi\sqrt{\frac{l}{g}}
\end{equation}

In order to present a more realistic expression for the period in the case of larger angles, it was proposed another approximated formula, derived from the complete elliptic integral of first kind \cite{kidd,millet}:
\begin{equation}
T\approx 2\pi\sqrt{\frac{l}{g\cos\theta_0}}
\end{equation}
where $\theta_0$ is the initial angular displacement and the amplitude of oscillation, supposing the the pendulum starts at rest.

In fact, series approximations are common in classical mechanics
textbooks \cite{fowles}, and another approximations were recently
proposed \cite{belendez1, belendez2}.

It is well known that the exact formula for the large-angle period of a simple
pendulum, involving the elliptic integral of the first kind
$K(\cdot)$\footnote{$K(\cdot)$ is implemented in most computer algebra
systems available in the market.} is given by \cite{drazin}:
\begin{equation}
\label{periodK}
T=4\sqrt{\frac{l}{g}}K\left( \sin^2{\frac{\theta_0}{2}}\right).
\end{equation}
Bel\'endez \emph{et al} \cite{belendez}, in a very didactic note, have derived this formula from the integral expression of $\theta(t)$. 

In this brief work we will essentially solve the same problem
treated by Bel\'endez \emph{et al} \cite{belendez}, but considering
a non-zero initial angular velocity. We present a closed-form
expression for the angular displacement $\theta(t)$, in terms of the
Jacobi elliptic function $\text{sn}(u,k)$, and of the incomplete
elliptical integral of the first kind $F(\varphi,k)$. We also made
some energy considerations, and derive an expression for the period
of oscillation $T(\theta_0,\phi_0)$, involving $K(\cdot)$. We
believed that such an exercise can be fruitfully used in order to
introduce such concepts as elliptical integrals and functions, to
encourage the use of Computer Algebra Systems (Maple$^{\circledR}$,
Mathematica$^{\circledR}$, etc.), and to show how a simple
generalization can be carried out.

\section{The Model}
The dynamics of an ideal simple pendulum is given by the following initial value problem:
\begin{equation}
\left\{
\begin{split}
&\ddot{\theta}+\omega_0^2 \sin \theta =0\\
&\theta(0)=\theta_0\\
&\dot{\theta}(0)=\phi_0
\end{split}
\right.
\end{equation}
where $\omega_0^2=\sqrt{\frac{g}{l}}$ is its natural angular frequency, $\theta_0\in[0,2\pi]$ is the initial angular displacement, and $\phi_0\in \Re$ is the initial angular velocity.

Following \cite{belendez}, we start to solve this second order differential equation, multiplying it by $\frac{d\theta}{dt}$:
\[
\frac{d\theta}{dt}\frac{d^2\theta}{dt^2}+\frac{d\theta}{dt}\omega_0^2 \sin \theta =0
\]
and noting that it can be rewritten as:
\[
\frac{d}{dt}\left[ \frac{1}{2} \left( \frac{d\theta}{dt} \right)^2 - \omega_0^2 \cos \theta \right] =0
\]
Then, integrating this equation on the time interval $[0,t]$, $t>0$, we obtain:
\[
\begin{split}
&\Rightarrow \left[ \frac{1}{2} \left( \frac{d\theta}{dt} \right)^2 - \omega_0^2 \cos \theta \right]_{0}^{t} =0\\
&\Rightarrow \frac{1}{2} \left( \frac{d\theta}{dt} \right)^2 -\frac{1}{2} \left( \frac{d\theta}{dt}\Big{|}_{t=0} \right)^2 = \omega_0^2( \cos \theta(t)-\cos \theta_0)\\
&\Rightarrow \left( \frac{d\theta}{dt} \right)^2=2\omega_0^2( \cos \theta(t)-\cos \theta_0)+\phi_0^2\\
\end{split}
\]
Now, the objective is to write this equation in order that, when we integrate it, we will obtain an elliptic integral. To do this, we define the new variables:
\begin{equation}
\label{change_var}
\begin{split}
&y(t)=\sin \left(\frac{\theta(t)}{2} \right)\\
&k=\sin^2 \left(\frac{\theta_0}{2} \right)
\end{split}
\end{equation}
and rewrite the equation as:
\begin{equation}
\label{dydt}
\left( \frac{dy}{dt} \right)^2=\omega_0^2 k\left(1-\frac{y^2}{k}\right)(1-y^2)+\frac{\phi^2}{4}(1-y^2)
\end{equation}
where we have used that
\[
\frac{dy}{dt}= \frac{dy}{d\theta} \frac{d\theta}{dt}=\frac{1}{2} \cos \left(\frac{\theta}{2}\right)\frac{d\theta}{dt}
\]
and that
\[
\cos^2\left(\frac{\theta}{2}\right)=1-\sin^2\left(\frac{\theta}{2}\right)
\]
Defining
\[
\tau=\omega_0 t
\]
\[
z=\frac{y}{\sqrt{k}}
\]
follows that:
\[
\frac{dz}{d\tau}=\frac{dz}{dy} \frac{dy}{dt} \frac{dt}{d\tau}=\frac{1}{\omega_0 \sqrt{k}}\frac{dy}{dt}
\]
Substituting $\frac{dy}{dt}$ on ($\ref{dydt}$) yields:
\begin{equation}
\label{dzdtau}
\begin{split}
\left( \frac{dz}{d\tau} \right)^2&= \left(1-z^2\right)(1-kz^2)+\frac{\phi_0^2}{4\omega_0^2 k}(1-kz^2)\\
&=(1-kz^2)(\gamma_0^2-z^2)
\end{split}
\end{equation}
where $\gamma^2_0=1+\frac{\phi_0^2}{4\omega_0^2 k}$, $k\in(0,1)$, and the transformed initial conditions are:
\[
z(0)=1
\]
\[
\frac{dz}{d\tau}\Big{|}_{\tau=0}=\frac{\phi_0}{2\omega_0}\cot \left(\frac{\theta_0}{2}\right)
\]
Taking the square root on both sides of ($\ref{dzdtau}$), and after some algebra, we have that
\[
d\tau=\pm \frac{dz}{\sqrt{(1-kz^2)(\gamma_0^2-z^2)}}
\]
Integrating on the interval $[0,\tau]$ we obtain:
\[
\tau=\pm \frac{1}{\sqrt{k}}\int_1^z\frac{dz}{\sqrt{\left(\frac{1}{k}-z^2\right)(\gamma_0^2-z^2)}}
\]
which is equivalent to:
\[
\pm \tau=\frac{1}{\sqrt{k}}\left[\int_0^z\frac{dz}{\sqrt{\left(\frac{1}{k}-z^2\right)(\gamma_0^2-z^2)}}-\int_0^1\frac{dz}{\sqrt{\left(\frac{1}{k}-z^2\right)(\gamma_0^2-z^2)}}\right]
\]
Now, we can write these two integrals in terms of the inverse Jacobi elliptic function and of the \emph{incomplete} elliptical integral of the first kind (Formula 219.00, pp. 58, \cite{byrd}):
\begin{equation}
\label{period0}
\pm \tau=\text{sn}^{-1}\left(\frac{z}{\gamma_0},\gamma_0^2 k\right)-F\left(\arcsin \left(\frac{1}{\gamma_0}\right),\gamma_0^2 k\right)
\end{equation}
\[
\Leftrightarrow \text{sn}^{-1}\left(\frac{z}{\gamma_0},\gamma_0^2 k\right)=F\left(\arcsin \left(\frac{1}{\gamma_0}\right),\gamma_0^2 k\right)\pm \tau
\]
Applying the Jacobi elliptic function $\text{sn}(\cdot,\cdot)$ on both sides of this equation
\[
\frac{z}{\gamma_0}=\text{sn}\left( F\left(\arcsin \left(\frac{1}{\gamma_0}\right),\gamma_0^2 k\right)\pm \tau,\gamma_0^2 k\right)
\]
and coming back to the original variables
\[
\sin\left(\frac{\theta}{2}\right)= \gamma_0 \sqrt{k}\;\text{sn}\left( F\left(\arcsin \left(\frac{1}{\gamma_0}\right),\gamma_0^2 k\right)\pm \tau,\gamma_0^2 k\right)
\]
we finally obtain the solution for the angular displacement as a function of time:
\begin{equation}
\label{solution}
\theta(t)=2\arcsin\left[ \gamma_0 \sqrt{k}\;\text{sn}\left( F\left(\arcsin \left(\frac{1}{\gamma_0}\right),\gamma_0^2 k\right)\pm \omega_0 t,\gamma_0^2 k\right)\right]
\end{equation}
where
\begin{equation}
\label{defs}
\gamma^2_0=1+\frac{\phi_0^2}{4\omega_0^2 k},\; k=\sin^2\left( \frac{\theta_0}{2} \right),\text{ and }\omega_0^2=\sqrt{\frac{g}{l}}.
\end{equation}
Observe that expression ($\ref{solution}$) represents two possible solutions: there are a plus and a minus signal in front of $\omega_0 t$. Then, we will choose the "positive branch" if the initial angular velocity is positive ($\phi_0>0$), and the "negative branch" if the initial angular velocity is non-positive ($\phi_0 \leq 0$).

In particular, $\phi_0=0\;\Rightarrow\; \gamma_0^2=1$, and ($\ref{solution}$) reduces to:
\[
\theta(t)=2\arcsin\left[ \sqrt{k}\;\text{sn}\left( F\left(\frac{\pi}{2}, k\right)-\omega_0 t, k\right)\right]
\]
From Formula 110.06 (pp. 9, \cite{byrd}), the following equality holds:
\[
F\left(\frac{\pi}{2}, k\right)=K(k)
\]
where $K(\cdot)$ is the \emph{complete} elliptical integral of the first kind. In this way we recover the solution obtained by \cite{belendez} for the case of a null initial angular velocity:
\begin{equation}
\label{before}
\theta(t)=2\arcsin\left[ \sqrt{k}\;\text{sn}\left( K(k)-\omega_0 t, k\right)\right]
\end{equation}

In Figure 1 we plot the solution ($\ref{solution}-\ref{before}$) for $\theta_0=\pi/2$, using the software Mathematica$^{\circledR}$, and in Figure 2 the corresponding angular velocity $\phi(t)=\dot{\theta}(t)$. The parameters used were: $\theta_0=\frac{\pi}{2}\text{ rd}$, $l=1\text{ m}$, and $g=9.8\text{ m/s$^2$}$ (these same values for $l$ and $g$ will be used in Figures 3 and 4 below).

\section{Some Energy Considerations}
The problem that we are analysing is conservative, because the
pendulum is only under the influence of Earth's gravitational field,
that is, we are not considering any kind of dissipative forces.

It is straightforward to show that the kinetic energy in the system at time $t$ is given by:
\[
K(t)=\frac{1}{2}ml^2\phi^2(t)
\]
where $\phi(t)=\dot{\theta}(t)$. The potential energy, considering the lowest level of the pendulum as the reference point, is:
\[
U(t)=mgl(1-\cos(\theta(t)))
\]
Then, the sum of these two quantities is constant over time, and equals the mechanical energy, $E$, of the system:
\[
E=mgl(1-\cos(\theta(t)))+\frac{1}{2}ml^2\phi^2(t)
\]
In particular, we know the initial angular displacement ($\theta_0$) and velocity ($\phi_0$). Consequently, the mechanical energy is given by:
\[
E=mgl(1-\cos(\theta_0))+\frac{1}{2}ml^2\phi_0^2
\]
and the following expression must be true for all time:
\[
gl(1-\cos(\theta(t)))+\frac{1}{2}l^2\phi^2(t)=gl(1-\cos(\theta_0))+\frac{1}{2}l^2\phi_0^2.
\]
Dividing this equation by $l^2$ and multiplying by $2$, we can
rewrite it as
\begin{equation}
\label{energy}
2\omega_0^2(1-\cos(\theta(t)))+\phi^2(t)=2\omega_0^2(1-\cos(\theta_0))+\phi_0^2(t)
\end{equation}

From ($\ref{energy}$), the angular velocity of the pendulum for the some displacement $\theta(t)$ is given by:
\begin{equation}
\label{energy1}
\phi(t)=\pm \sqrt{2\omega_0^2(\cos(\theta(t))-\cos(\theta_0))+\phi_0^2}
\end{equation}

\subsection{Stopping at the top}

For instance, if we want to know what initial angular velocity $\phi_0$ is necessary, given that we know $\theta_0$, to the pendulum stops at its maximum height, we need only to consider $\theta(t)=\pi$ and $\phi(t)=0$ in ($\ref{energy1}$), obtaining:
\begin{equation}
\label{energy2}
\phi_0=\phi_c:=\pm \omega_0\sqrt{2(1+\cos(\theta_0))}
\end{equation}
In conclusion: if $|\phi_0|<|\phi_c|$, the pendulum will oscillate forever around its equilibrium point (its lowest level); if $|\phi_0|=|\phi_c|$, the pendulum stops at its maximum height; and if $|\phi_0|>|\phi_c|$ the pendulum will execute a circular motion in relation to its fixing point. In Figure 3 we show a graphic with the relation $|\phi_c|$ versus $\theta_0$.

\subsection{Pendulum's Period for $|\phi_0|<|\phi_c|$}
For $\phi_0=0$, remember that the pendulum's period is given by ($\ref{periodK}$):
\begin{equation}
T=\frac{4}{\omega_0}K\left( \sin^2\left(\frac{\theta_0}{2}\right) \right) = \frac{2}{\pi}T_0K(k)
\end{equation}
where $T_0=2\pi\sqrt{\frac{l}{g}}$ is the period of the pendulum for small oscillations ($\ref{periodsmall}$).

If $\phi_0\neq0$, we can still use this formula, just replacing
$\theta_0$ by $\theta_{max}$, that is, the maximum angular position
where the pendulum stops. We can obtain this value setting
$\phi(t)=0$ in ($\ref{energy1}$), and isolating
$\theta=\theta_{max}$:
\begin{equation}
\label{thetamax}
\theta_{max}=\arccos \left( \cos(\theta_0)-\frac{1}{2}\phi_0^2\right)
\end{equation}
Then, the period of the pendulum, for any initial position and velocity (considering $|\phi_0|<|\phi_c|$), is given by:
\begin{equation}
T(\theta_0,\phi_0)=\frac{2}{\pi}T_0K\left( \sin^2\left(\frac{\arccos \left( \cos(\theta_0)-\frac{1}{2}\phi_0^2\right) }{2}\right)\right)
\end{equation}
In Figure 4 we plot $T(\theta_0,\phi_0)$ and verify that the pendulum's period is an increasing functions of both $\theta_0$ and $\phi_0$.

\section{Concluding Remarks}
In this exercise we have derived a closed-form solution for the
angular displacement of a simple pendulum in terms of the Jacobi
elliptic function $\text{sn}(u,k)$, and of the incomplete elliptical
integral of the first kind $F(\varphi,k)$:
\[
\theta(t)=2\arcsin\left[ \gamma_0 \sqrt{k}\;\text{sn}\left(
F\left(\arcsin \left(\frac{1}{\gamma_0}\right),\gamma_0^2
k\right)\pm \omega_0 t,\gamma_0^2 k\right)\right]
\]
where
\[
\gamma^2_0=1+\frac{\phi_0^2}{4\omega_0^2 k},\; k=\sin^2\left(
\frac{\theta_0}{2} \right),\text{ and
}\omega_0^2=\sqrt{\frac{g}{l}}.
\]
and $\theta_0$, $\phi_0$ are the initial angular displacement and
velocity, respectively.

In addition, we have shown that, for an initial angular velocity
satisfying $|\phi_0| < \omega_0\sqrt{2(1+\cos(\theta_0))}$, the
pendulum will oscillate with a period equals:
\[
T(\theta_0,\phi_0)=\frac{2}{\pi}T_0K\left( \sin^2\left(\frac{\arccos
\left( \cos(\theta_0)-\frac{1}{2}\phi_0^2\right) }{2}\right)\right)
\]

Finally, such an exercise may be used to initiate advanced
undergraduate students to concepts such as elliptical integrals and
functions, to the use of Computer Algebra Systems
(Maple$^{\circledR}$, Mathematica$^{\circledR}$, etc.), and to show
how a simple generalization can be carried out.

\newpage

\begin{center}
\begin{figure}[h!]
\epsfig{file=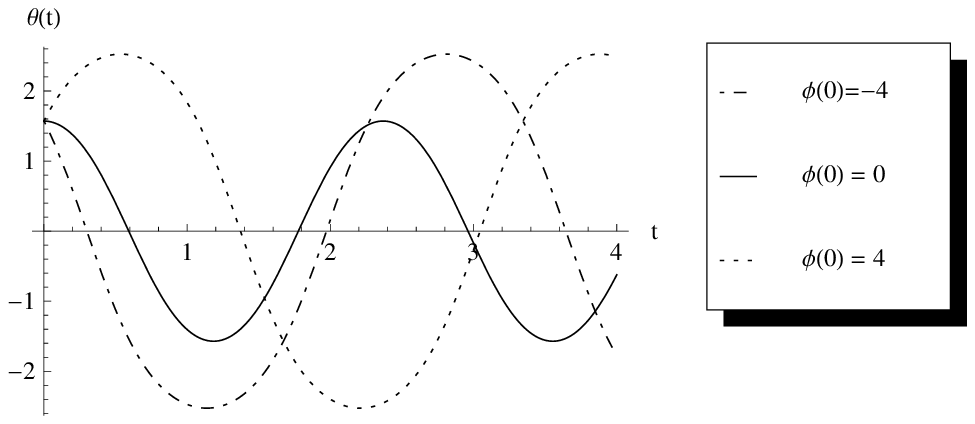,width=13cm,clip=}
\caption{Angular Position \emph{versus} Time}
\end{figure}
\end{center}

\begin{center}
\begin{figure}[h!]
\epsfig{file=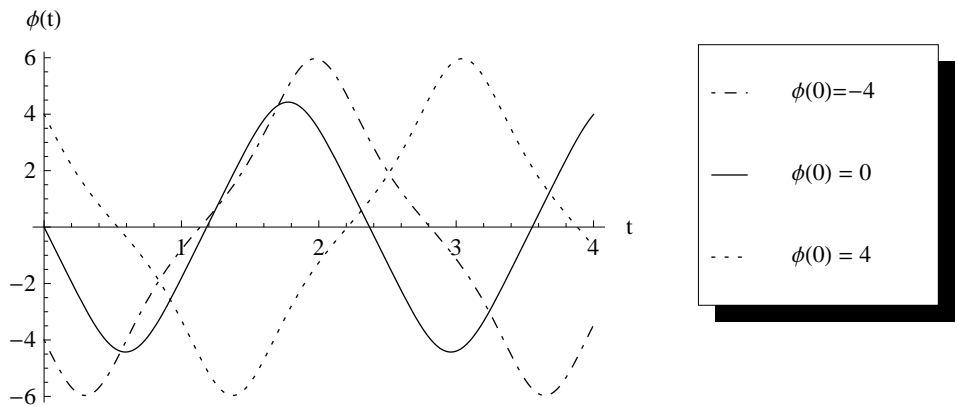,width=13cm,clip=}
\caption{Angular Velocity \emph{versus} Time}
\end{figure}
\end{center}

\begin{center}
\begin{figure}[h!]
\epsfig{file=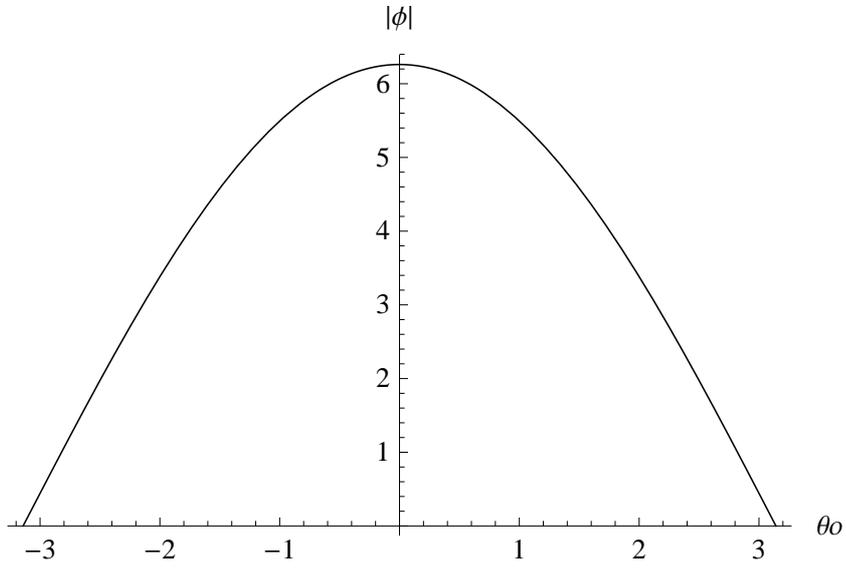,width=11cm,clip=}
\caption{$|\phi_c|$ \emph{versus} $\theta_0$}
\end{figure}
\end{center}

\begin{center}
\begin{figure}[h!]
\epsfig{file=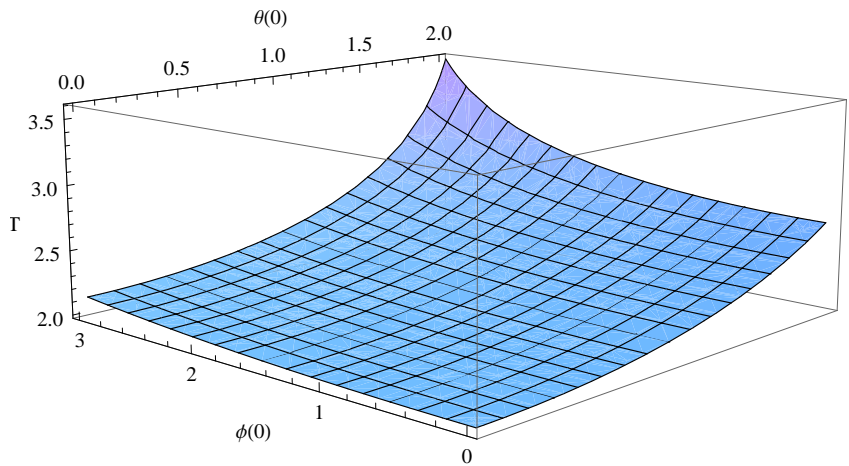,width=11cm,clip=}
\caption{$T(\theta_0,\phi_0)$}
\end{figure}
\end{center}

\end{document}